\documentclass{iopconfser}
\usepackage{a4wide,amssymb,graphics}
\usepackage{epsfig}
\usepackage{hyperref}
\begin{document}

\noindent Conference Proceedings for BCVSPIN 2024: Particle Physics and Cosmology in the Himalayas\\Kathmandu, Nepal, December 9-13, 2024 

\title{Neutrino Non-Standard Interaction and Implications}

\author{Anjan Giri}
\affil{Department of Physics, IIT Hyderabad, Kandi - 502284, India}
\email{giria@phy.iith.ac.in}

\begin{abstract}

In this talk, we discuss the possible implications of physics beyond the standard model in the neutrino sector applying non-standard interaction. Recently, the NOvA and T2K neutrino experiments have announced their measurements for the CP violating parameter $\delta_{CP}$. The observed value of T2K is around 1.5$\pi$, while NOvA gives the same parameter value around 0.8$\pi$ in the normal mass order scenario. Although it is too early to make any conclusions about the difference in measurements but attempts have been made in the literature to invoke the new physics ideas to explain the difference. We include the dual non-standard neutrino interaction effects and find interesting results. Furthermore, we show the possibility of seeing an appreciable difference in the normal and inverted neutrino mass hierarchies if one measures the CP asymmetries in the upcoming long-baseline neutrino experiments.
\end{abstract}

\section{Introduction}

Understanding of Neutrino physics has come a long way since Pauli proposed the existence of neutrinos in 1930. The most elusive particles in nature, "Neutrinos", are present in abundance in the Universe. These neutral particles do not feel strong interactions and interact via weak interactions. Neutrino Physics has undergone tremendous development in the last few decades, with many interesting results and, therefore, new experiments planned for the future. At this stage, the nature of neutrino and many of its properties are unknown. The popular belief is that once the two upcoming long-baseline neutrino experiments are in operation we will be able to come to know many aspects of neutrinos and may eventually move to the precision era in neutrino physics. Observation of Neutrino Oscillation confirmed that Neutrinos are massive particles, but we are yet to determine the mass of neutrinos. The discovery of Neutrino oscillation ~\cite{Super-Kamiokande:1998kpq} paved the way for increased scrutiny from both theory and experimental fronts. Six oscillation parameters determine the neutrino oscillation probabilities in the SM with three neutrinos: three mixing angles ($\theta_{12},\theta_{13},\theta_{23}$), one Dirac CP phase $\delta_{CP}$, and two mass-squared differences ($\Delta m^{2}_{21}, \Delta m^{2}_{31}$). While efforts are on to directly measure its mass, oscillation probability gives us the mass square differences which give rise to the so-called mass hierarchies in the neutrino sector, namely, the normal mass hierarchy and the inverted mass hierarchy. In the scenario of normal mass hierarchy, the mass eigen states are such that the first one being the lowest mass and the third one is the highest mass (assuming 3 flavour neutrino formalism), whereas in the inverted hierarchy scenario, the third one is the lightest among the three states [Ref....]. Interestingly, this is one of the goals of the upcoming neutrino experiments, in addition to the measurement of CP parameter $\delta_{CP}$ and other parameters. We hope to learn soon, out of the two possibilities, which of these possibilities nature actually prefers. Since already we got an indication of non-zero neutrino mass from the observed neutrino oscillation experiments (in the standard model neutrinos are massless particles) it is widely believed that Neutrino oscillation is an ideal avenue to probe new physics beyond the standard model. 

In our Universe we have more matter than antimatter, which is known as the matter-antimatter asymmetry (also known as the baryon asymmetry of the Universe (BAU)). Combined operation of Charge and Parity symmetry (CP) is found to be violated in the weak interaction. Moreover, CP violation has already been observed in the bottom quark meson (B) meson systems but the observed CP violation in the quark sector will not be able to explain the observed BAU and hence additional CP violation needed from elsewhere. In this context, CP violation from the neutrino sector is likely to help, if observed. It must be stressed here that with the measurement of $\theta_{13} \ne 0$ the stage is cleared for the measurement of CP violation in neutrino sector.
One can show that at least around 2 orders of magnitude of additional contribution can be obtained from the neutrino CP (assuming maximal CP violation). Apart from the contribution from the leptonic sector, additional sources of CP violation beyond the standard model may also be needed to help explain the observed data. 

Looking at the current scenario, there are three important challenges that exist in the neutrino sector: the value of CP-violating phase $\delta_{CP}$, $\theta_{23}$ octant degeneracy, and the sign of mass-squared difference ($\Delta m^{2}_{31}$), which are likely to be known in the upcoming neutrino experiments. The neutrino oscillation programs have been able to measure the unknown oscillation parameters with increasing accuracy.

The neutrinos, while traveling through the Earth's crust, get influenced by a matter potential known as Wolfenstein's matter effect~\cite{Wolfenstein:1977ue}. Wolfenstein introduced non-standard interaction (NSI)~\cite{Grossman:1995wx} besides the neutrino mass matrix to probe new physics. Neutrino phenomenology has been extensively studied in the literature. NSI describes the non-standard neutrino interaction with ordinary matter, which is parameterized by effective couplings, $\epsilon_{\alpha \beta}$, where $\alpha$ and $\beta$ are the generation indices. The standard neutrino oscillation in the matter can get affected by NSI, and determining these subdominant effects will be the major role of the upcoming long-baseline neutrino experiments. Needless to mention that there have been lot of quark related to neutrino NSI in the literature and NSI effect can pollute the clean determination of the parameters, in particular the $\delta_{CP}$. 

 Recently, T2K and NO$\nu$A  announced their results on neutrino CP-violating parameters. While for the inverted mass ordering, they look consistent with each other, but for the normal mass ordering, there appears to be some tension (T2K points to the $\delta_{CP}$ about 1.5$\pi$ whereas NO$\nu$A suggests the same parameter to be around 0.8$\pi$)~\cite{NOvA:2021nfi, T2K:2021xwb}. It will be intriguing to learn more about neutrino mass ordering from the results of the ongoing experiments T2K-NO$\nu$A, and it is expected to be a persistent issue in neutrino physics that will be resolved in the DUNE-T2HK era. Interestingly, attempts have been made to link the difference in $\delta_{CP}$, observed in T2K and NO$\nu$A, to the possibility of new physics in the form of non-standard interactions~\cite{Chatterjee:2020kkm, Denton:2020uda, Brahma:2022xld}. It may be noted here that there has been joint effort by both NO$\nu$A and T2K experiments to provide combined fit results and we hope to get better clarity in the near future. 

 It has been shown that introduction of NSI effect help alleviate the difference in the CP measurements by the two experiments. Here, we attempt to understand the CP-violation through the CP asymmetry parameter in the presence of non-standard neutrino interactions arising simultaneously from $\epsilon_{e\mu}$ and $\epsilon_{e\tau}$ sectors, by considering the results of T2K-NO$\nu$A ~\cite{T2K:2021xwb, NOvA:2021nfi}. We focus on utilizing the NSI constraints in the CP asymmetry study with varying energy. These NSI constraints are obtained by scanning two different NSI coupling parameters $\epsilon_{e\mu}$ and $\epsilon_{e\tau}$ at a time. The possible implications of CP asymmetry w.r.t. varying energy could help us decipher interesting details on neutrino mass ordering. The results discussed here are for both DUNE and T2HK experimental setups.




\subsection{Neutrino oscillation in the presence of NSIs}

The NSI is defined by dimension six four-fermion ($ff$) operators of the form~\cite{Wolfenstein:1977u}
\begin{equation}\label{1}
    {\mathcal{L}}_{NSI} = 2\sqrt{2}G_{F} \epsilon_{\alpha \beta}^{fC} [  \overline{\nu_{\alpha}} \gamma^{\rho} P_{L}  \nu_{\beta}][\overline{f} \gamma_{\rho} P_{C} f] + h.c.
\end{equation}
In the above expression, $\epsilon_{\alpha \beta}^{fC}$ is the dimensionless parameters that assess the strength of the new interaction with respect to the SM, the three SM neutrino flavor $e, \mu, \tau$ are denoted by $\alpha$ and  $\beta$, matter fermions $f$ by  $u, d, e $, and the superscript $C = L, R$ refers to the chirality of $ff$ current. 

The matter potential $V$ in the presence of NSI is written as $V = 2\sqrt{2}G_{F}N_{e}E$. 
For neutrino propagation in Earth, the NSI coupling: \begin{center}
    $\epsilon_{\alpha\beta}e^{i\phi_{\alpha \beta}} \equiv \sum_{f, C}\epsilon_{\alpha\beta}^{fC} \frac{N_{f}}{N_{e}} \equiv \sum_{f=e,u,d}(\epsilon_{\alpha\beta}^{fL}+\epsilon_{\alpha\beta}^{fR}) \frac{N_{f}}{N_{e}}$,
\end{center}
$N_f$ being the number density of $f$ fermion and $N_{e}$ the number density of electrons. For this analysis, we utilise flavor non-diagonal NSI ($\epsilon_{\alpha \beta}$'s with $\alpha \neq \beta$). Mainly, we focus on the dual NSI parameter $\epsilon_{e \mu}$ and $\epsilon_{e \tau}$ (arising simultaneously) to examine the conversion probability of $\nu_{\mu}$ to $\nu_{e}$. For dual NSI scenario, the oscillation probability expression for $\nu_{\mu} \rightarrow \nu_{e}$ can be written as the sum of four (plus higher order; cubic and beyond) terms: 

\begin{equation}
P_{\mu e} = P_{SM} + P_{\epsilon_{e\mu}} + P_{\epsilon_{e\tau}} + P_{Int} + {h.o.}
\end{equation} 
where, the higher-order terms of the NSI parameters. $P_{SM}$ is the SM contribution, $P_{\epsilon_{e\mu}} $$, P_{\epsilon_{e\tau}}$, ${\rm and}$~ $P_{Int}$ are terms containing respective NSI contributions and interference terms between $\epsilon_{e\mu}$ and $\epsilon_{e\tau}$ contributions. 

Using the oscillation probability, we study the CP asymmetry observable. The CP asymmetry is the difference between the quantity of matter and anti-matter in the universe. The CP asymmetry measures the change in oscillation probabilities when the sign of the CP phase changes. CP-asymmetry is defined as:
\begin{equation}
    A_{CP} \equiv \frac{P(\nu_{\alpha}\rightarrow\nu_{\beta})-P(\overline{\nu}_{\alpha}\rightarrow\overline{\nu}_{\beta})}{P(\nu_{\alpha}\rightarrow\nu_{\beta})+P(\overline{\nu}_{\alpha}\rightarrow\overline{\nu}_{\beta})},
\end{equation}

Here, along with $A_{CP}$ observable, we use another observable for our illustration.
\begin{equation}
    \Delta A_{\alpha \beta}(\delta_{CP}) \equiv A_{\alpha \beta}(\delta_{CP}\neq0) - A_{\alpha \beta}(\delta_{CP}=0),
\end{equation}  
\section{Analysis Details}
For our analysis purpose, we use the software GLoBES and associated tools. The standard model parameters best-fit values are taken from NuFIT v5.2. In particular, the parameter values taken for our analysis (for normal ordering) are: $\sin^{2}\theta_{12}=  0.303^{+0.012}_{-0.012}$;\hspace*{0.1cm}$\sin^{2}\theta_{13}= 0.02225^{+0.00056}_{-0.00059}$;\hspace*{0.1cm}
$\sin^{2}\theta_{23}=0.451^{+0.019}_{-0.016}$;\hspace*{0.1cm} $\delta_{CP}=232^{+36}_{-26}$;
$\frac{\Delta m^{2}_{21}}{10^{-5}eV^{2}}= 7.41^{+0.21}_{-0.20}$; and $\frac{\Delta m^{2}_{3l}}{10^{-3} eV^{2}}=+2.507^{+0.026}_{-0.027}$. As mentioned earlier, we assume that the new physics effects are responsible for different values of $\delta_{CP}$, and in this case, it is due to the presence of NSIs. 

We use the available data from NO$\nu$A and T2K experiments to constrain the dual NSI parameters and the associated new phases as shown in Fig. 1, consistent with the recent analysis \cite{nova:2024abc}. Thereafter, we have consider the effect of NSI on LBL experiments, DUNE, T2HK, and a combination of DUNE and T2HK. We have shown how in the presence of dual NSI contributions how the oscillation probabilities differ as in Fig. 2. One can see from the figure that while measuring the neutrino and anti-neutrino probabilities one can see significant difference, which may be tested in the future experimental facilities.

In Fig. 2 (top panel), the oscillation probability plots for DUNE and T2HK in neutrino mode in the SM (left panel), SM with dual NSI arising simultaneously from the $e-\mu$ and the $e-\tau$ sector (right panel) are shown. We see a good separation between NO-IO for both $\delta_{CP}=90^\circ$ and $\delta_{CP}=-90^\circ$ in the SM scenario. For the SM with dual NSI scenario, we still have some separation between NO-IO for $\delta_{CP}=90^\circ$ in the mid-energy region. Whereas in the case of $\delta_{CP}=-90^\circ$, the NO-IO separation continuously decreases, and they gradually merge around 4 GeV. For the anti-neutrino scenario, we see a reasonable separation between NO-IO for $\delta_{CP}=90^\circ$, 
 and $\delta_{CP}=-90^\circ$ for both SM and SM with dual NSI case.

\begin{figure}[htbp]
\minipage{0.45\textwidth}
\includegraphics[width=\linewidth]{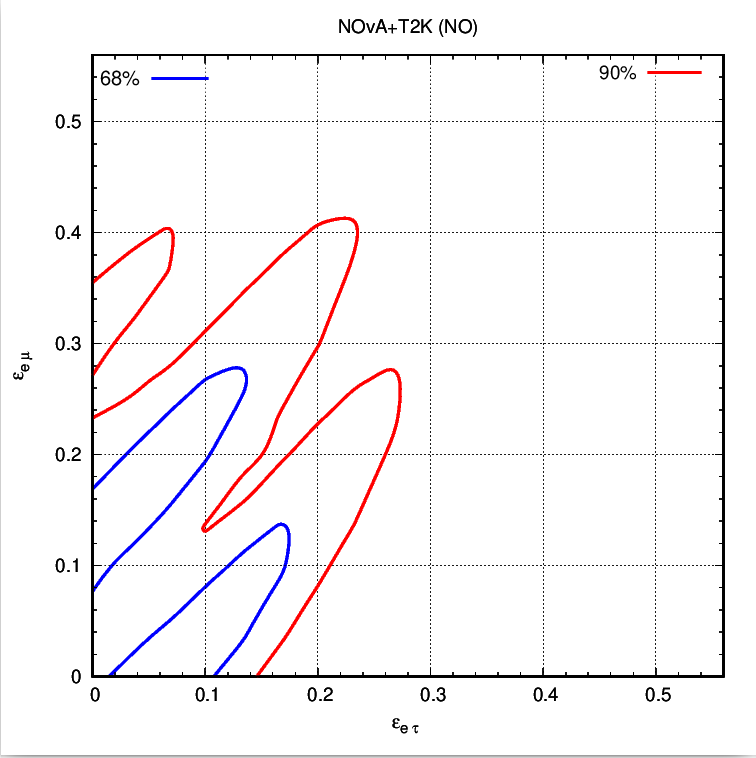}
\endminipage\hfill
\minipage{0.45\textwidth}
\includegraphics[width=\linewidth]{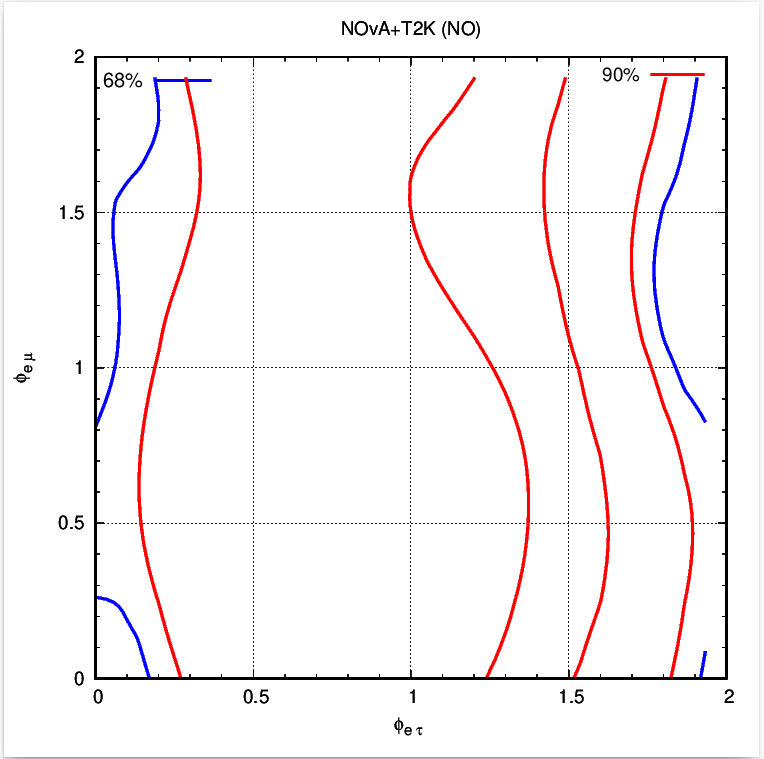}
\endminipage
\caption{Probability Plots for T2HK in SM (left) and SM with dual NSI scenario with NSI arising from both $e-\mu$ sector and $e-\tau$ sector (right) for $\nu$ (top panel) and $\bar{\nu}$ (bottom panel) mode}\label{fig:image9}
\end{figure} 
\begin{figure}[htbp]
\minipage{0.45\textwidth}
\includegraphics[width=\linewidth]{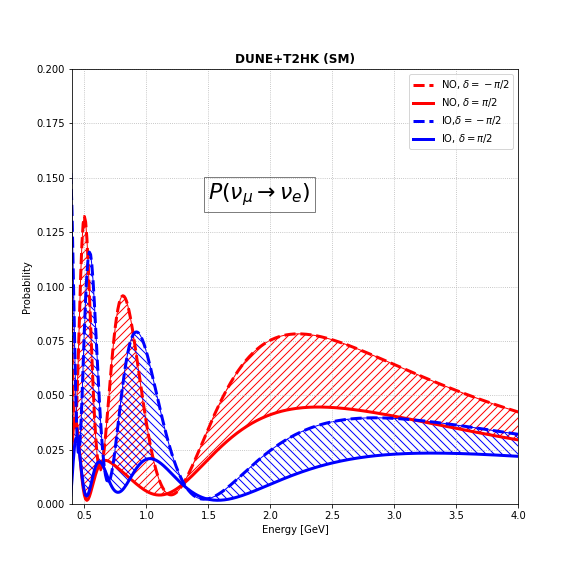}
\endminipage\hfill
\minipage{0.45\textwidth}
\includegraphics[width=\linewidth]{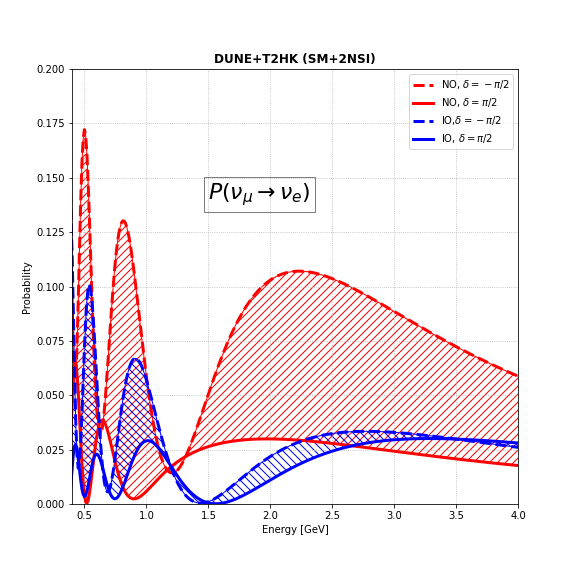}
\endminipage
\caption{Probability Plots for DUNE+T2HK in SM (left) and SM with dual NSI scenario with NSI arising from both $e-\mu$ sector and $e-\tau$ sector (right) for $\nu$ (top panels)}\label{fig:image10}
\end{figure}

\begin{figure}
\minipage{0.44\textwidth}
\includegraphics[width=7.2cm,height=7.0cm]{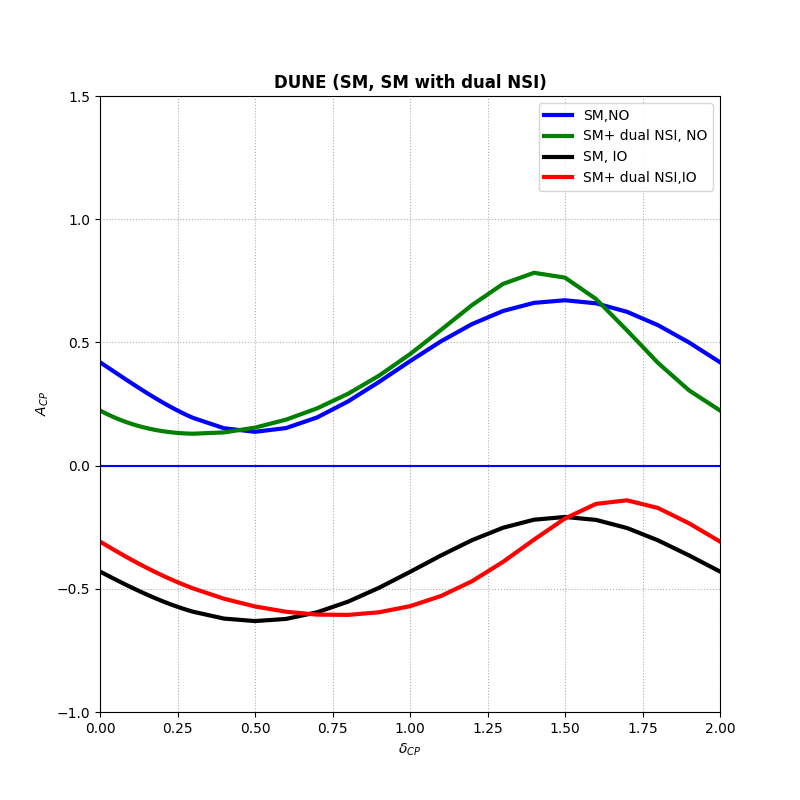}
\endminipage\hfill
\minipage{0.44\textwidth}
\includegraphics[width=7.2cm,height=7.0cm]{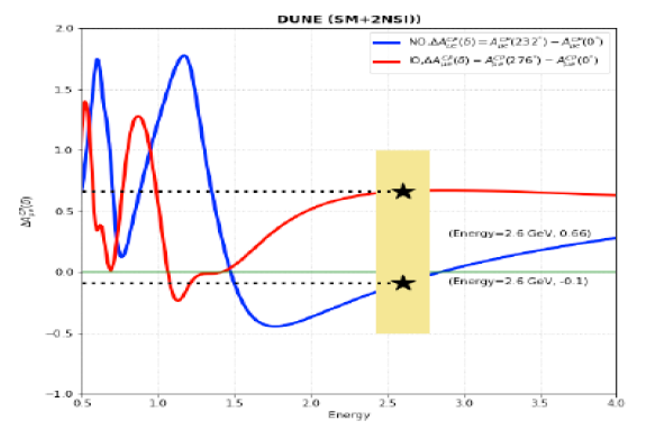}
\endminipage
\caption{ In the left plot the CP asymmetry for DUNE experiment has been shown for NO and IO including the dual NSI. In the right plot CP asymmetry for both NO and IO are presented for using the specific input points from the global fit\label{fig:image12}}
\end{figure}

\section{\bf{Remarks}}
In this talk, we considered the recent results from the two ongoing long baseline neutrino experiments, T2K and NOvA, in particular the $\delta_{CP}$ values. We assumed that new physics occurs in the form of NSI contributing simultaneously from $e-\mu$ and $e-\tau$ sectors. In doing so, we obtained the dual constraints on NSI parameters by combining the NO$\nu$A and T2K datasets. It may be noted that our NSI coefficients are realistic in the sense that we have derived them from the latest results from accelerator-based neutrino experiments and are consistent with latest results. We observed the striking effects of dual NSI constraints on both neutrino and anti-neutrino channel probabilities in DUNE, T2HK, and a combination of DUNE and T2HK. We observed that, this can help us to understand and possibly distinguish the neutrino mass ordering problem. Here, the study of CP asymmetry reveals that with the inclusion of dual NSI from $e-\mu$ and $e-\tau$ sectors, simultaneously, the CP asymmetry provide very much different results to get a hint of the normal or inverted mass hierarchy in the DUNE experiment due to its long baseline in comparison to the T2HK scenario. As is known presence of NSI can act a spoiler for the clean determination of the neutrino parameters but more studies may also help us to get a better handle on the mass hierarchies problem in the long baseline experiments.      


\large{Acknowledgement: We acknowledge the support from the DST, India.}

\end{document}